\documentstyle[12pt]{article}
\begin{document}
\input psbox.tex

\def\sqr#1#2{{\vcenter{\vbox{\hrule height.#2pt
          \hbox{\vrule width.#2pt height#1pt \kern#1pt
           \vrule width.#2pt}
           \hrule height.#2pt}}}}
\def\square{\mathchoice\sqr68\sqr68\sqr{4.2}6\sqr{3}6}

\newcommand{\figprov}[4]{
\begin{figure}[hbt]
\begin{center}\mbox{\psboxto(#1;0cm){#4}}\end{center}
\caption{\it #3}
\label{#2}
\end{figure}
}

\begin{flushright}
ULB-TH 95/06\\
UMH-MG 95/02\\
LPTENS 95/18\\
May 1995\\
\end{flushright}
\vskip 1 truecm
\centerline{\bf{Hawking Radiation Without Transplanckian Frequencies.
}}
\vskip 1. truecm
\centerline{R. Brout\footnote{e-mail: smassar @ ulb.ac.be},\addtocounter{footnote}{-1} 
S. Massar\footnotemark $\/^{,}$\footnote{Boursier IISN},
}
\centerline{Service de Physique Th\'eorique, Universit\'e Libre de Bruxelles,}
\centerline{Campus Plaine, C.P. 225, Bd du Triomphe, B-1050 Brussels, Belgium}
\vskip 5 truemm
\centerline{R. Parentani\footnote{Unit\'e propre de recherche du C.N.R.S.
associee a l' l'Ecole
Normale Sup\'erieure et a l'Universite de Paris Sud. 
e-mail: parenta@physique.ens.fr}}
\centerline{Laboratoire de Physique Th\'eorique de l'Ecole 
Normale Sup\'erieure} 
\centerline{24 rue Lhomond
75.231 Paris CEDEX 05, France}
\vskip 5 truemm
\centerline{Ph. Spindel\footnote{e-mail: spindel @ umh.ac.be}}
\centerline{M\'ecanique  et Gravitation, Universit\'e de Mons-Hainaut,}
\centerline{Facult\'e des
Sciences, 15 avenue Maistriau, B-7000 Mons, Belgium}
\vfill 
\setcounter{footnote}{0}

\newpage
{\bf Abstract }In a recent 
work,
Unruh showed that 
Hawking radiation is unaffected by a truncation of free field theory at
the Planck scale. 
His analysis was performed numerically and based on a hydrodynamical 
model.
In this work, by analytical methods, 
 the mathematical and physical origin of
Unruh's result is revealed. An alternative truncation scheme which may
be more   appropriate
for black hole physics is proposed and analyzed. In both
schemes the thermal Hawking radiation remains unaffected even though
transplanckian energies no  longer appear. The universality
of this result is explained by working in momentum space. 
In that representation, in the presence of a horizon, the d'Alembertian
equation 
becomes a
singular first order equation. In addition, the boundary conditions
corresponding to vacuum before the black hole formed are that the in--modes
contain positive momenta only. Both properties remain  valid when 
the spectrum
is truncated and they suffice to obtain Hawking radiation.
\vfill \newpage

\section{Introduction}\label{SS1}

The theory of black hole evaporation \cite{Hawk} is sore beset with two unsolved
dilemmas: \begin{itemize}
\item[1)] The transplanckian issue \cite{THooft}\cite{Jacobson1}: when deriving
Hawking radiation in the usual framework of free field theory, one calls upon
vacuum fluctuations at $ {\cal I}^-$ \cite{EMP}\cite{MaPa} 
whose energies are $ O (e^{M^2} / M)$.
These propagate as such up to a Planckian distance of 
the horizon where there
energy is redshifted down to $O(1)$ and then further 
down to a typical frequency of $ O (M^{-1})$ upon reaching
$ {\cal
I}^+$.
Since gravitational interactions become strong at the
Planckian scale free field theory is at best dubious.

\item[2)] The unitary issue \cite{Hawk3}:  in the semi-classical theory of back
reaction, both the matter which is the source of gravity (the star), and the
``partners'' to the emitted Hawking photons fall into the singularity, giving
rise to a density matrix description of the radiation, thus, in the last
analysis, to a non-unitary description of the evolution. Can one incorporate
such a quandary into quantum mechanics?

\end{itemize}

\noindent Both will require a deeper knowledge of how gravity reacts to Hawking
emission in order to be resolved. But perhaps at a more preliminary stage
progress may be made by introducing an effective theory which one guesses
incorporates some of the features which might emerge from the fundamental
theory. Such is the nature of a recent interesting contribution of W.
Unruh \cite{Un94}, who addressed himself to the taming of the transplanckian monster.
The present paper, inspired by Unruh's work, contains an analysis as well as a
generalization of the taming mechanism.

Unruh's work is based on an analogy between the hydrodynamic equations of motion
of sound waves in a moving fluid and those that governs s-wave emission of a
massless scalar field from an incipient Schwarzschild black hole in free field
theory. Thus one is led to predict the production of a thermal flux of phonons
as the fluid passes from sonic to supersonic flow \cite{Un81}. Through numerical
computation he subsequently showed \cite{Un94} that a truncation of the
spectrum of sound -using a rather natural algorithm- in no way affected
the thermal emission of Hawking phonons in the circumstances that this
emission occurred in the untruncated theory. Carrying this lesson over
to the black hole situation, the implication is that tinkering with the
transplanckian part of the photon spectrum  will leave the thermal
emission unaffected. We shall here follow up on Unruh's tinkering, first
by introducing a truncation scheme which we believe to be more
appropriate to the black hole situation, and then
 supply
the mathematical rationale for the resistance of Hawking's result to
such mutilation. Throughout we have preserved the linearity of the
field equations. Whether non linear effects preserve Hawking radiation
remains a moot point.

The paper is organized in seven parts. Section \ref{SS2} contains a brief review
of Unruh's considerations. In preparation for our analysis we present in Section
\ref{SS3} a variant of the technique of Damour and Ruffini 
 \cite{DaRu}, using
momentum space considerations. This technique provides for a simple and elegant
characterization of initial conditions which then rapidly leads to the
understanding which we seek. In Section \ref{SS4}, the photon spectrum is
truncated in this approach. It follows immediately from the formalism that
tinkering with the transplanckian part of the spectrum does not affect the
Hawking thermal emission. Section \ref{SS5} incorporates the Damour-Ruffini technique into the
Unruh hydrodynamic truncated model. Using similar reasoning as in Section
\ref{SS4}, the usual thermal emission is once more recovered. Section \ref{SS6}
contains a comparison of the wave packet trajectories in the two cases that have
been analyzed so as to draw a physical picture of the production mechanisms.
Finally in Section \ref{SS7} we speculate on the physics behind our
phenomenological truncation procedure. Some interpretation in terms of quantum
gravity is hazarded. The result of our analysis can be viewed as one of
disappointment. Hawking radiation does not seems to be, in itself,  an open door
that leads to quantum gravity. Rather, it provokes thought in that direction,
without offering, at least in a direct way, an orientation for the solutions.

\section{Unruh's Hydrodynamic Model}\label{SS2}

The equation that governs the propagation of sound (amplitude $ \equiv \phi $) 
in a perfect fluid in 1 + 1 dimensions which flows with a stationary background
velocity  field $V(\xi)$ is

\begin{equation}
\label{E1}
[ \left( \partial_{\eta} + \partial_\xi V (\xi) \right) \;
(\partial_{\eta} + V (\xi) \partial_{\xi} ) - \partial_{\xi} ^2 ] \phi =
0 \qquad .  
\end{equation}

\noindent
This equation can be rewritten in d'Alembertian form  $ \square \phi =
0 $ with metric $ g^{00} = 1, \; g ^{01} = g^{10} = V, \; g^{11} = - 1 + V^2$.
It is readily diagonalized by the transformation $ dt = d \eta + V d\xi / (1 -
V^2) \; ; \; d \xi = d r$ to read

\begin{equation}
\label{E2}
\square \phi = [ ( 1 - V^2)^{-1} \partial_t ^2 - \partial_r ( 1 - V^2)
\partial_r ] \; \phi = 0 \qquad . 
\end{equation}

In this form one recognizes the s-wave part of $ \square$ constructed from the
Schwarzschild metric with the identifications: $ t, r= $ Schwarzschild time and
radial coordinate respectively; $ V^2 = 2 M / r$. The horizon is at $ \vert V
\vert = 1$. Thus a fluid, whose flow rate approaches $ V = - 1$, will emit a
thermal distribution of phonons whose temperature is given by $ (1/2 \pi) d V /
d r \vert _{V = -1}$. 
At this point we simply use the form given by  eq. (\ref{E1})
to motivate, with Unruh \cite{Un94}, 
the truncation procedure used for this model.

In the rest frame $ (V = 0)$, the spectrum of sound determined from 
 eq. (\ref{E1}) is $ \omega = \vert k \vert $. But since the fluid has atomic
structure, the spectrum has the property $ \omega = \vert k \vert $ when $ k
< k_0$ and $ \omega \to \omega_0$ for $ k >> k_0$. 
 For example $ \omega = k_0$ tanh $(k/k_0)$ might be expected to
approximate the spectrum. In general for $ \omega = F (k)$, we are then led to
modify  eq. (\ref{E1}) to

\begin{equation}
\label{E3}
[ \left( -i\partial_{\eta} - i\partial_{\xi} V \right) \left(
-i\partial_{\eta} - V i\partial_{\xi} \right) - F^2
\left(-i\partial_{\xi}\right) ] \phi = 0 \qquad .
\end{equation}

This equation has been the starting point of Unruh's numerical computations.
Propagating backwards in time an outgoing wave packet centered about a given
negative frequency, he determines the Bogoljubov coefficients by decomposing
the packet at early times into its positive and negative frequency components.
Setting up a vacuum state at early times (See Section \ref{SS5}), the $ \beta$
term in the Bogoljubov transformation (i.e. the weight of the positive
frequency part at early times of a negative frequency mode at late times)
encodes the presence of outgoing phonons. This latter conforms to the
existence of the outgoing thermal Hawking flux - quite independently of the
truncation function, $F$.

An amusing aside is provided by Landau's theory of superfluid critical
velocity, $ V_s$. Were vortices and rotons absent, the approach to critical
superfluid flow would be accompanied by a thermal shower of Hawking phonons.

\section{The Damour - Ruffini Method}\label{SS3}

To explain and generalize Unruh's result we shall use the Damour-Ruffini 
technique \cite{DaRu}, using a
variant based on momentum space introduced in ref. \cite{PaBr0}.
It is especially convenient for our present purpose because the wave equation
near the horizon and the boundary conditions defining in-modes take a
particularly simple form.

Equation~(\ref{E2}) is valid outside a collapsing spherical star. First we
transform it through use of the advanced Eddington-Finkelstein system, $v$
and $r$, where $ v = t + r^*, r^* = r + 2 M\ell n [ (r - 2 M) / 2 M ]$ to give

\begin{equation}
\label{E4}
\square \psi = - [ \partial_ r (1 - 2 M /r) \partial_r + 2
\partial_v \partial_r ] \psi (v,r) = 0
\end{equation}

\noindent The solutions, $ \psi$, are connected smoothly to those which are
inside the star. Since the details of the star's trajectory are irrelevant in
the interesting asymptotic region where the star's surface approaches the
horizon ($r = 2 M$), we shall for simplicity of presentation take a very simple
model: the star is defined by a thin shell radially falling inward with the
speed of light. We choose its trajectory to be $ v = v_{st} = 0$ where $st$
labels the star's surface. 

Inside the star, for $v<0$, the geometry is flat and
described by the metric $ ds^2 = d v^2 - 2 d v d r$ ( In terms of usual
Minkowski time $T$ one has $ v = T + r).$ In this coordinate system waves obey 

\begin{equation}
\label{E5}
\square {\psi} = -\partial_r (\partial_r + 2 \partial_v) \psi = 0
\end{equation}

\noindent 
The usual Minkowski modes inside the star hereafter called in-modes, are

\begin{eqnarray}
\chi_{\lambda} ^{in (v)} ( v)  &=& e^{-i \lambda v} / \sqrt{4 \pi \lambda}
\qquad ,\label{E6} \\
\chi_{\lambda} ^{in (u)} (v, r) &=& e^{-i \lambda (v - 2 r)} / \sqrt{4 \pi
\lambda}\qquad .\label{E7}
\end{eqnarray}

\noindent The $u$-sector of the field operator is $ \hat \phi = \int_{0}
^{\infty} d \lambda \left( a_\lambda \chi_\lambda^{in (u)} + a_\lambda^\dagger
\chi_\lambda^{in (u) *} \right) $ i.e. a superposition of in-modes given by
 eq. (\ref{E7}) inside the star. The Heisenberg state is the vacuum state 
$\vert{0_{in }}\rangle$ which is annihilated by the destruction operators $ a_\lambda$,
i.e. there are no quanta inside the star. For a full description one should
consider $v$-modes $ \chi_\lambda ^{(v)}$ as well. But these do not encode
particle creation and will not be considered here. Henceforth we drop the label
$(u)$. (To be complete we also point out that in order to describe the s-wave
sector of a 3 + 1 dimensional theory one should impose the additional boundary
condition that the modes vanish at $r = 0$. This complicates the notations
without modifying   the result and will not  be taken  into account.)

 Hawking radiation is encoded in the history of the outgoing modes $
\chi_\lambda^{in}$ as they propagate through the star into the space exterior
to it. In particular we shall display their form outside the star both near and
far from the horizon. In the first instance, the equation that governs their
behavior is

\begin{equation}
\label{E8}
-\partial_x [ ( x / 2 M) \partial_x + 2 \partial_v ] \psi (x, v) = 0
\end{equation}

\noindent where $ x \equiv r - 2 M$. This equation holds in the domain 
$\vert x \omega \vert <<1$
where $\omega$ is the eigenvalue of $i \partial_v$. Indeed
the phase of the exact solution of eq. (\ref{E4}) $e^{-i \omega
v}e^{-i 2 \omega ( x+ 2M \ln x)}$ and the phase of the
solution of  eq. (\ref{E8})  $e^{-i \omega v}e^{-i  \omega 4M \ln
x}$ differ by $O(1)$ when $\vert x \omega \vert =O(1)$.
Henceforth we shall consider only typical $\omega =
O(M^{-1})$ which implies that $x<M$.

\par From eq. (\ref{E6}), the 
$v$-modes (irrelevant for production) remain of the form $ e^{-i \omega v} / \sqrt{4 \pi \omega}$, whereas the
$u$-modes are given by the linear combination

\begin{equation}
\label{E9}
{\cal F}_\omega (x, v) = { e^{-i \omega v}\over \sqrt{4 \pi \omega} }[ \theta
(x)
 A x^{4 M i \omega} + \theta (- x) B \vert x \vert^{4 M i \omega}] 
\end{equation}

\noindent Standard Klein Gordon normalization prescribes $ \vert A \vert^2 -
\vert B \vert^2 = 1$.

 Far from the horizon $ (x \gg 2 M)$, the wave equation reduces to

\begin{equation}
\label{E10}
- \partial_x \left( \partial_x + 2 \partial_v \right) \psi
=0
\end{equation}

\noindent and the $u$-modes $ \psi_\omega^{out}$ associated to asymptotic quanta
on $  {\cal I} ^+$ are equal to $ \psi _\omega ^{out} = e^{- i \omega ( v - 2
r)} / \sqrt{4 \pi \omega  }$. Since the
exact solutions of eq. (\ref{E4}) which connects to this asymptotic form is $
\psi_\omega ^{out} = e^{- i \omega (v - 2 r^*)} / \sqrt{4 \pi \omega}$, the
out-modes near the horizon are given by $ A = 1, \; B = 0$ in eq. (\ref{E9}):

\begin{equation}
\label{E11}
\psi_\omega ^{out} = {1 \over  \sqrt{4 \pi \omega}} e^{-i \omega v} 
\theta
(x) x^{ 4 M  i \omega}\qquad .  
\end{equation}

As the crux of our analysis lies in a careful formulation of the matching
conditions at the star's surface which define  the modes outside in terms of
the modes inside the star  we now go into this
matter in some detail.
In order to determine the out-particles content of the in-vacuum, it is 
propitious to compute the Fourier transform of the modes 
at the surface of the star ($ v  = 0$). The matching conditions are
then implemented in simple and elegant fashion.

The Fourier transform of  eq. (\ref{E8}) is

\begin{equation}
\label{E12}
[ p \partial_p +  4 M i \omega + 1  ] {\tilde {\cal F}}_\omega (p) = 0
\end{equation}

\noindent whence

\begin{equation}
\label{E13}
{\tilde{\cal F}}_\omega (p) = {\sqrt{M} \over \sqrt{2} \pi}{1 \over p } \; [ C
p^{-4 M i \omega} \theta (p) + D \vert p \vert^{-4 M i \omega} \theta (-p) ]
\qquad .
\end{equation}

\noindent The domain of validity of eqs. \ref{E12}) and \ref{E13}) is $ \vert p
\vert \gg   M^{-1}$ (for typical $\omega = O(M^{-1})$).

The Fourier transform of the Minkowski modes  eq. (\ref{E7}) are proportional 
to $ \delta (p
- 2 \lambda)$, and we 
recall that positive norm modes have $ \lambda > 0$.

 The object of the exercise is to use continuity to change the basis
from modes inside the star $ \left( \equiv \tilde \chi_\lambda^{in}(p)\right)$
to those outside the star which for large $p$ are of the from $ {\tilde {\cal
F}} _\omega (p)$. Since the former set have $ p > 0$ (since $ \lambda > 0)$,
continuity prescribes that $ p > 0$ is valid for the latter set as well. In
this way one establishes that an equivalent set of in modes is found from Eq.
(\ref{E13}) with $ D = 0$ and $C=1$ (this set was obtained independently 
from Damour-Ruffini, but by working in Kruskal coordinates by
Unruh \cite{Unru1} and Hawking \cite{Hawk3}). We call this bases $ {\tilde
\psi}_{\omega}^{in} (p)$

\begin{equation}
\label{E14}
{\tilde {\psi}}_\omega ^{in} (p) = {\sqrt{M} \over \sqrt{2} \pi}
 \theta (p)  p^{-4
M i \omega - 1} \qquad .
\end{equation}

\noindent This gives $ {\tilde{\psi}}_\omega ^{in} (p)$ for sufficiently large
$p$ (e.g. $ p \gg M^{-1}$). Thus for these
values of $p$ one finds that the expansion coefficients $ \alpha _{\omega
\lambda}$ which give $ {\tilde{\psi}}_ \omega ^{in} (p)$ in terms of $ {\tilde
{\chi}}_\lambda^{in} (p)$ are $ (M/ 8 \pi^2 \lambda)^{1/2} ( 2 \lambda)^{- 4 M
i \omega}$. 
For smaller values
of $p$ which are relevant for the construction of packets
which cross the surface of the star at large values of $x$, as we have stated
above, $ {\tilde{\psi}}_\omega^{in} (p)$ and $ {\tilde{\chi}}_\lambda^{in}
(p)$ coincide. In this case $ \alpha_{\omega \lambda} \to \delta (\omega -
\lambda)$.

To see explicitly that only the large values of $p$ are relevant for
packets which reach ${\cal I}^+$ at late times, one 
builds a wave packet with the modes eq. (\ref{E14}):
\begin{equation}
\tilde \varphi^{in}_{\omega_0, u_0}(v,p) 
=
\int\! d \omega \ e^{i\omega u_0}{ e^{-(\omega - \omega_0)^2/ 2 \sigma^2}
 \over ({2 \pi})^{1/4} \sigma^{1/2}}
e^{-i \omega v} {\tilde
{\psi}}_\omega ^{in} (p)
\label{E15B}
\end{equation}
The phase $e^{i\omega u_0}$ centers the wave packet on the light ray
$v-2r^* = u_0$ and the gaussian $
e^{-(\omega - \omega_0)^2/ 2 \sigma^2}$ centers the frequency around
$\omega_0$. Inserting the form of ${\tilde
{\psi}}_\omega ^{in} $ one obtains
\begin{equation}
\tilde \varphi^{in}_{\omega_0, u_0} (v,p)
=
{\sqrt{M} ({2 \pi})^{1/4} \sigma^{1/2}\over \sqrt{2} \pi}
{1 \over p}
e^{-(v + 4M \ln p -u_0)^2 \sigma^2/2}
e^{i \omega_0(v + 4M \ln p -u_0)}
\label{E15C}
\end{equation}
whereupon it is seen that the wave packet is centered on the trajectory
$v + 4M \ln p =u_0$. The momentum $p$ at the surface of the star $v=0$
is given by $p= e^{u_0/4M}$. Hence for wave packets which reach ${\cal
I}^+$ at late times $u_0$,  $p$ at the surface of the star is
much larger than $M^{-1}$.

In conclusion, we have the important result that, for all values of $p$, only
positive ones appear in the basis functions $ {\tilde{\psi}}_\omega^{in} (p)$.
Furthermore, for $ p \gg M^{-1}$ which are those necessary to
describe outgoing packets which begin their journey from the surface of the
star to $ {\cal I}^+$ at values of $  x_{st} \ll M$, the content of the $
{\tilde{\chi}}_\lambda^{in} (p)$ modes in terms of $ {\tilde{\psi}}_\omega ^{in}
(p)$ contain both signs of $ \omega$. On the contrary for $ p \to  2 \omega$,
which are the relevant values to describe outgoing packets which begin their
journey to $ {\cal I}^+$ at values of $  x_{st} \gg M$, $\omega $ becomes equal
to $\lambda$. Therefore it is the former set that gives rise to particle 
creation
on  $ {\cal I}^+$, whereas the latter give rise to no creation. In this
formalism, this is what expresses the well known fact that Hawking radiation
sets in as the star's surface approaches the horizon exponentially closely. To
derive these results all that has been used is continuity at the star's
surface. From now on we shall characterize in-vacuum by $ a_\omega \vert
{0_{in}} \rangle= 0$ where $ a_\omega$ are the annihilation operators
associated with the $ \psi_\omega^{in}$ modes and we re-emphasize that $
\omega$ takes on both signs in this characterization.

To find the content of $ \vert{ 0_{in}}\rangle$ in terms of the out-modes (defined on $
{\cal I}^+)$ one considers the Fourier transform of $ {\tilde{\psi}}^{in} (p)$
valid for $ p \gg M^{-1}$, hence $ x \ll M$

\begin{eqnarray}
\psi_\omega^{in} (x) &=& {\sqrt{M} \over \sqrt{2} \pi}
 \int^{\infty}_{0} {d p } \; e^{i p x}
\; p^{-4 M i \omega - 1} \hspace{4cm} \nonumber\\
&=& {\sqrt{M} \over \sqrt{2} \pi}{\Gamma (-4 M i \omega) } \vert x  \vert^{4 M i
\omega} [ \theta (x) e^{+ 2 \pi M \omega} + \theta (-x) e^{-2 \pi M \omega} ]
\quad . \label {E15} 
\end{eqnarray}

\noindent The mode $ \psi_\omega ^{in}$ lies on both sides of the horizon. Only
the piece outside the horizon (i.e. proportional to $ \theta (+x))$ propagates
out to $ {\cal I}^+$. From eqs. (\ref{E11}) and (\ref{E15}) one has outside the
horizon $ (x > 0)$

\begin{eqnarray}
 \psi_\omega ^{in} (x) \theta (+ x) = \left\{ \begin{array}{ll}
\alpha_{\omega} \psi_\omega^{out} (x) \quad \mbox{ $\omega
> 0$}\\
\beta_{\vert \omega \vert} \psi_{\vert
\omega \vert}^{out *}(x) \quad \mbox{$\omega < 0$}
\end{array}
\right.   \label {E16}
\end{eqnarray}

\noindent 
where we have introduced the Bogoljubov coefficients

\begin{eqnarray}
\alpha _\omega &=& {\Gamma (- 4 M i \omega) \sqrt{2 M \omega}\; e^{2 \pi M \omega}
\over \sqrt{ \pi}} \qquad ,\nonumber \\
\beta_\omega &=& {\Gamma (+4 M i \omega) \sqrt{2 M \omega}\; e^{-2 \pi M
\omega} \over \sqrt { \pi}}\qquad ,   \nonumber\\
\vert \beta_\omega / \alpha_\omega \vert &=& e^{-4 \pi M \omega} \qquad .
\label {E17}
\end{eqnarray}

\noindent This implies a steady 
thermal flux of emitted particles at temperature $ T_H
= 1 / 8 \pi M$.

 The
condition $p>0$ used in  eq. (\ref{E14}) 
in the present formalism is equivalent
to the Damour-Ruffini requirement that $ \psi_\omega^{in} (x)$ be analytic in the upper
half $x$ plane. What we have seen here is that it is a direct consequence of the
existence of vacuum in the star. The description we have given here is
readily adapted to take into account truncation of the transplanckian spectrum
$(p > 1)$.

\section{Confronting the Transplanckien Hiatus}\label{SS4}

The mechanism of how Hawking photons get created is through the combined
gravitational and Doppler red shift which is incurred as a wave packet voyages
from small values of $x$ to $ {\cal I}^+$. In the steady state of radiation the
star's surface is exponentially close to the horizon $ (x_{st} \simeq M e^{-
t/2M})$. In consequence the $p$ values, obtained from  eq. (\ref{E15}), which
dominate the integrand in the packet in this region are exponentially large:
the saddle point of the integrand in  eq. (\ref{E15}) is at $ p^* = 4 M \omega /
x$ \cite{PaBr0}. 

To see this more precisely, recall that the locus of saddle points traces out
the classical trajectories of photons. Near $x = 0$ they are obtained from the
Hamiltonian
constraint 

\begin{eqnarray}
H \equiv p \left( {xp \over 2 M} + 2 p_v \right) (=0) \label {E18}
\end{eqnarray}

\noindent where we have used Eddington-Finkelstein coordinates $x,v$ and their
conjugate momentum $p$ and $p_v$ (compare with  eq. (\ref{E8})).
The canonical equation  are 
\begin{equation}
 {\dot p} = - \partial H / \partial x\ ;\ 
{\dot x} = \partial H / \partial p\ ;\ {\dot v} = \partial H / \partial 
\omega\ ;\ {\dot p_v} =0\quad , 
\end{equation}
where the dot denotes derivative with respect to an affine parameter along the
trajectory. These combine to yield 
\begin{eqnarray}
p_v&=&-\omega=const. \qquad ,\\
{dp \over dv}& =& -{p\over 4M}\qquad . \label{E20a}
\end{eqnarray}
 On mass
shell, $H$ vanishes and accordingly
\begin{eqnarray}
x = {4 M \omega \over p} \qquad .\label {E19}
\end{eqnarray}
\noindent Hence

\begin{equation}
p(v) = p_{st} e^{- v/4M} \quad \mbox{ and} \quad x(v) 
= {4 M \omega \over p_{st}}
e^{v/4M} \label {E20}
\end{equation}

\noindent where $p_{st}$ is a  constant of integration. From this last equation,
we see that a photon of fixed energy $\omega$, erected at late time and
reaching $ x = O (M)$ at $ v_*$ (where $v_*$ is typically of order the life
time of the black hole $=O(M^3)$) must  have crossed the surface of
the star  at $ x_{st}  = 4 M \omega /
p_{st} =e^{-v_*/4M} O (M) $ with an enormous momentum $ p_{st} = 
e^{+ v_*/ 4M} O(\omega)$. This is the transplanckian problem. At these high
values of $p$, and concommitantly small values of $x$, free field theory
certainly breaks down. The gravitational interaction of the s-wave modes under
consideration, both with other modes and with the background field whose source
is in the degrees of freedom of the star becomes enormous. Therefore the
underpinnings of the free field calculation become completely fallacious.

This does not mean, however, that the derivation given in Section \ref{SS3} is
completely lost. This is Unruh's main point. We shall first illustrate these
considerations in the context of the Damour-Ruffini formalism of Section \ref{SS3} by
introducing a truncation in that scheme and in Section \ref{SS5} proceed to
analyze Unruh's effective theory as given in Section \ref{SS2}. In both of
these schemes one explores the hypothesis that the physics near the horizon can
be mimicked by modifying the way the matter field propagates, but leaving both
gravity and the linear character of the fluctuations unaffected.

First a few words of clarification are in order. Any truncation scheme can be
formulated in intrinsic geometric terms. However it is convenient to work in a
coordinate system which is privileged in the geometry of the incipient black
hole. We make the assumption that the truncation takes a simple form in such a
privileged system. Explicitly, the origin is fixed and the angular momentum
expansion is carried out with respect to it. The radius has an intrinsic
geometric meaning. Furthermore the 
space time outside the star is static and the origin of
time is given by the inception of radial infall.

Let us therefore begin by truncating in the Eddington-Finkelstein system. We suppose that the
energy spectrum gets modified for modes probing space time at scales less then
1, i.e. for $ p > 1$.  Then  eq. (\ref{E8}) 
is assumed to be modified to the form
\begin{equation}
g(-i\partial_x) \left[
{x \over 2M} g(-i\partial_x) - 2 i \partial_v \right] \psi(x,v) =0\qquad .
\label{E21a}
\end{equation}
Hence the mode equation corresponding to Eq \ref{E12}) becomes

\begin{eqnarray}
\label{E21}
[ \partial_p  g(p) + 4 M i \omega  ] 
\tilde {\cal F}_ \omega (p) = 0\qquad .
\end{eqnarray}

\noindent The physics that goes into the evaluation of $ { g} (p)$ will be
the subject of the conjectural discussion of Section \ref{SS7}. For the present
we only require that $ { g} (p) = p$ for $ p < O(1)$ and $ { g} (p)
= 1$ for $ p > 1$ since we anticipate that it is only the Planckian and
transplanckian modes which will be affected by gravity. As in Unruh's model,
the function $ g(p)$ is now to be identified with the energy associated with
the mode number $p$ rather than $p$ itself. Indeed in flat space the proposed
modification  eq. (\ref{E21a}) reads 

\begin{eqnarray}
\label{E22}
g(-i\partial_x) [ g (-i\partial_x) - 2 i\partial_v] \psi = 0
\end{eqnarray}

\noindent so that one sees that for outgoing modes $ \lambda$,  the eigenvalue
of $ i\partial_v$, is equal to $g(p)/2$ rather than $p/2$ and is bounded by $1/2$.
In the curved space outside the star, the energy at $ {\cal I}^+ \  (\equiv
\omega)$ is Doppler and gravitationally red shifted when compared to the energy
measured by the free falling observer (for a
discussion of the red shift see ref \cite{GO} Chapter 3 and Appendix D). This
red shift is represented by the factor $ x/4 M$ in  eq. (\ref{E21a}) and the
energy measured by the free falling observer is $ 4M\omega /x = g(p)$ which
is now bounded by $1/2$. 

It is instructive to see how the truncation deforms the classical trajectories
from the geodesic associated with the free field. The
truncated version is described by the Hamiltonian

\begin{eqnarray}
\label{E23}
{\cal H} \equiv -{ g} (p) [ {x \over 2M} { g} (p)  +  2 p_v ] \ ( = 0 )
\end{eqnarray}

\noindent and eqs (\ref{E20a})  and (\ref{E19}) now read

\begin{eqnarray}
p_v&=&-\omega \qquad ,\\
d p/ dv &=& - { g} (p) / 4 M \qquad .\label{E24a}\\
x &=&  4 M \omega / g (p) \qquad ,\label {E24} 
\end{eqnarray}

\noindent Integrating, one sees that $\vert p \vert$ decreases with $v$, initially
(when $\vert p\vert >1$) linearly and then exponentially 
(when $ \vert p \vert <1$). For
$ \vert p \vert  >1$, one has $ \vert x \vert  = \vert 4 M \omega \vert$ and for 
$\vert p \vert <
1$, $\vert x \vert $ is proportional to $\vert p \vert^{-1}$, hence increases
exponentially with $v$. Thus at the same time as one avoids transplanckian
energies, one ceases to approach the horizon to within transplanckian distances
(typically $\omega M = O(1)$).
Section \ref{SS6} contains a sketch of the production process based on these
classical trajectories (see Fig. \ref{Ufigc}).

To establish Hawking radiation in the truncated theory we once more have to
characterize the in-modes exterior to the star. Subsequently it must be shown
that these in-modes evolve towards ${\cal I }^+$ so as to give the required
radiation.
\par From the truncated equation inside the star (eq. (\ref{E22})) it is seen
that  positive $\lambda$ 
implies positive $p$. Thus continuity across $v=0$
once more implies
that in-modes on the outside also have positive $p$. (This crucial
result, which is independent of the details of $g(p)$, results from our
assumption of postulating a linear field equation which keeps the
d'Alembertian factorized into a $u$ and $v$ piece. That this is sufficient is
obvious, but that it is not necessary will become apparent in Section
\ref{SS5}). Thus our in-modes, solution of  eq. (\ref{E21}) with $p>0$ are
\begin{equation}
\tilde \psi ^{in}_\omega (p) ={ \sqrt{M} \over \sqrt{2} \pi}\theta(p) g(p)^{-1}
e^{-4 i M \omega \int^p dp / g(p)}\qquad .
\label{E25}
\end{equation}
The Fourier transform of  $\psi ^{in}_\omega (x)$ has contributions from
$p >1$ for $x = O(1)$ and $p<1$ for $x>O(1)$. This is seen by
examining the saddle point condition in the Fourier transform which in fact
reconstructs the classical trajectories  eq. (\ref{E24}). The piece which is
relevant for Hawking production is at $x>>1$, hence concerned with cisplanckian
values of $p$ and  one has 
\begin{eqnarray}
\psi ^{in}_\omega (x)\vert_{\mbox{\rm cis}}
&\simeq& \int^{1}_{M^{-1}} dp { \sqrt{M} \over \sqrt{2} \pi}e^{ipx}
\tilde
\psi ^{in}_\omega (p)\nonumber\\ 
&=& \int_{M^{-1}}^1 dp { \sqrt{M} \over \sqrt{2}
\pi}e^{ipx} \vert p \vert ^{-4 M i \omega -1} \nonumber\\ 
&\simeq& \int^{\infty}_0 dp { \sqrt{M} \over \sqrt{2}
\pi}e^{ipx} \vert p \vert ^{-4 M i \omega -1} \quad ; \ x>>1 \label{E26}
\end{eqnarray}
thereby recovering  eq. (\ref{E15})\footnote{
We point out that eq. (\ref{E22}) has in general a large number of
linearly independent solutions (an ``infinity'' if $g(\partial_x)$ is not
polynomial in $\partial_x$). These linearly independent solutions are
recovered from the function $ \tilde
\psi ^{in}_\omega (p)$ by specifying which integration contour is used
to take the Fourier transform. According to how the singularities of $\tilde
\psi ^{in}_\omega (p)$ are encircled, different solutions are obtained.
We have chosen the contour to coincide with the path in the absence of
truncation, namely to lie on the real axis. Other prescriptions would
presumably give rise to runaway solutions and the properties of the
theory at low momentum would not coincide with the free field theory.}.

It should be noted that the norm of the cisplanckian modes is the same 
as in the
free field case. This results from the fact that the truncated problem is
possessed of a conserved norm $(= 2\pi\int \!dp \ 
\tilde \psi ^{*}_\omega (p)\  g(p)\ \tilde \psi_{\omega^\prime}(p)
= 
\delta(\omega - \omega^\prime))$ which reduces to the free field
form for $\vert p \vert <1$. Thus the decomposition of  
$\psi ^{in}_\omega (x)\vert_{\mbox{\rm cis}}$
into pieces localized at $x>0$ and $x<0$
 as in  eq. (\ref{E15}) has its usual interpretation
in terms of unitarity.

When translated into the variable $x$, what we have shown is that the free field
vacuum is maintained at $x\geq O(1)$ since one only requires the
characterization of the in-state in terms of its cisplanckian content in this
region. Most succinctly, Unruh vacuum is still a valid concept just outside a
Planckian skin of the horizon. This is every bit as valid as the statement that
Unruh vacuum is the correct description of the in-state in free field theory
outside the star. A little bit of Planckian fuzz around the horizon does no
injury to the physics since the conversion to Hawking photons on ${\cal I}^+$
occurs outside the star due to the redshift which the outgoing modes feel on the
scale of $x=O(M)$ and not $x=O(1)$ (this has been pointed out by
Jacobson \cite{Jacobson2} who however did not derive the insensitivity of
Hawking radiation to  planckian tinkering).

The picture that emerges is that fluctuations within this skin steadily develop
into outgoing pairs. Note that from  eq. (\ref{E24a}), $p$ in this
region grows linearly in $v$. The interpretation is that there is a conversion
of modes within the skin to become the usual outgoing modes of free field
theory, thereby guaranteeing a steady state. Thus at large values, $p$ has
become converted from an energy to a mode counting parameter. Note that the
total time of evaporation is $O(M^3)$, so that the total
number of modes which boil off from those initially inside the skin is $O(M^2)$
which is proportional to the usual estimate for the entropy of the black hole.

\section{The Truncated Unruh Model}\label{SS5}

In this section we analyze the production of Hawking phonons based on Unruh's
truncation  eq. (\ref{E3}). As in Section \ref{SS4}, the analysis is based on the
momentum representation of modes. Complications occur because the equation for
the modes near the horizon is second order and it is necessary to control that
the transplanckian sector does not contaminate the cisplanckian physics. This
was obvious in Section \ref{SS4} once the boundary conditions were set.

To have a clear idea of the mechanism of production from the modes, it is first
worthwhile to go into the classical motion along trajectories. Whilst this part
of the analysis does not give the production per se, it does give the
motion of wave packets before and after production. In this way one has
a guide to the portion of the mode analysis which is relevant.

The Hamiltonian, which generates the classical trajectories corresponding to
the wave equation  eq. (\ref{E3}) is

\begin{equation}
\label{E4.1}
H = \left [ p_\eta + p\/ V  (\xi) ] ^2 - F^2 (p) \right] \ (= 0)
\end{equation}

\noindent
In the context of  eq. (\ref{E4.1}), $p_\eta$ is the momentum conjugate to $
\eta$; $p$ being conjugate to $ \xi$. From the canonical equations  one finds

\begin{eqnarray}
p_\eta &=&-\omega=const.\qquad ,\\
d p / d \eta & =& - p V^{\prime}(\xi) \qquad , \label{E4.2}\\
d \xi / d \eta&=&\left \{ \begin{tabular}{ll}
$V(\xi)+F^\prime (p)$ on trajectories of type II or III\\
$V(\xi)-F^\prime (p)$ on trajectories of type I
\end{tabular}
\right.
\end{eqnarray}

\noindent Since the fluid flows to the left with $V$ decreasing to the left (so
that near the horizon $ V = - 1 + \xi / 4 M)$, it is seen that $ \vert p \vert$
always decreases in $\eta$. As a consequence we can, and shall, describe
evolution with respect to $ \eta$ in terms of $p$.

\figprov{10cm}{UFig1}{The two curves $-\omega \pm \vert F(p)\vert$  (solid
curves) and the family of lines $-V(\xi) p$ (doted lines) are plotted as
functions of $p$ for some representative values of $\xi$. Their intersections
give the trajectories $\xi(p)$. Their are three trajectories labeled I, II,
III. The set I corresponds to a $v=const$ trajectory. The set II crosses the
horizon when $\xi =0$ and then starts to propagate for $\xi < O(-4 M
\omega)$ whereupon it corresponds to a $u=const$ trajectory. The points in
class III never reach the horizon: there are no solutions in this class
for $\xi < \xi_{min}(\omega)=O(4 M \omega)$ but their are two solutions for
$\xi < \xi_{min}$ corresponding to a trajectory which bounces. 
}{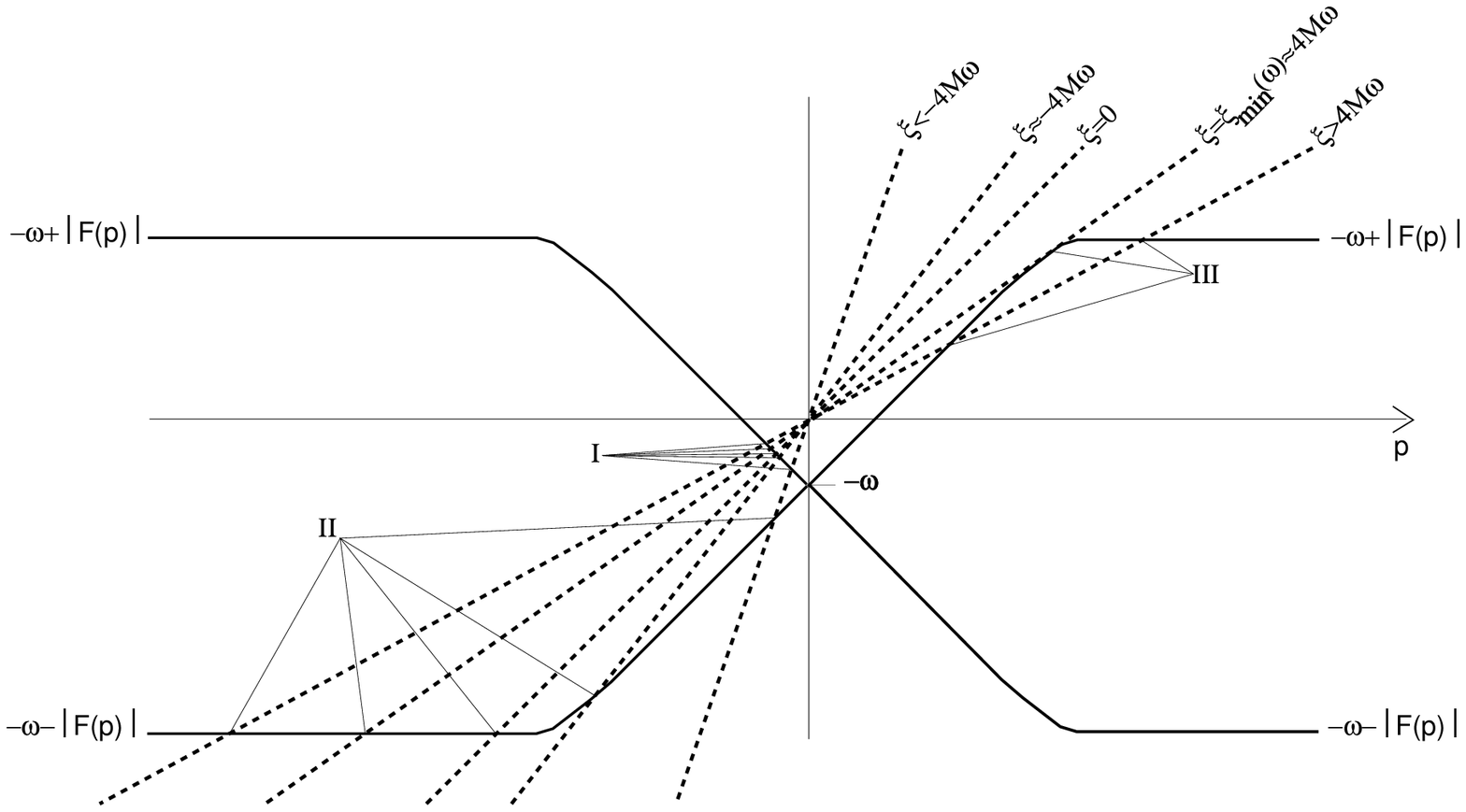}

To analyze the trajectories we use the on mass shell condition
$-pV(\xi) = -\omega \pm \vert F(p)\vert$. In Fig.
\ref{UFig1} is plotted as functions of $p$, the family of curves $-V
(\xi) p$, and the two curves $ -\omega \pm \vert F(p) \vert $. The intersections
give the trajectories $ \xi (p)$ at fixed $ \omega$. It is seen that there are
three types of trajectories:

\begin{itemize}
\item[I] The sequence of points near the origin all have small negative
$p$, hence correspond to a usual null geodesic in Schwarzschild geometry which,
in E F coordinates is $v$= constant. They are therefore uninteresting for the
production of particles. 

\item[II] These are points which, for $ \xi \gg - 4 M \omega$, have $ p \ll 1$,
hence $ F(p)$= 1. A particle on this part of the trajectory is at rest
with respect to the fluid. For $ \xi < - 4 M \omega$, $p$ is $<$ 1 and the
particles then propagate with respect to the fluid. This trajectory stops at $
p = 0$ owing to the monotonous character of $ p(\eta)$ ( eq. (\ref {E4.2})).

\item[III] The third trajectory has $ p > 0$. Once more for large positive $p$
it is non propagating. As $ \vert p \vert$ decreases, it starts to propagate for
$ p = O(1)$ whereupon it reaches a minimum value of $ \xi ( \equiv \xi_{min}
(\omega)) $ which is $ O ( 4 M \omega)$. As $ \vert p \vert$ continues to
decrease, $ \xi$ now increases so as to describe the propagating part of the
trajectory for negative values of $p$. 
\end{itemize}

Trajectories II and III are plotted in Fig.\ref{Ufigc} where the functions $
\xi (v)$ are displayed (these are very similar to the functions $\xi(\eta)$
since the
coordinate transformation used to go from  eq. (\ref{E1}) to  eq. (\ref{E2}), shows
that near the horizon $ \eta$ differs from $v$ by a regular function of $\xi$.).
Production is concerned with the mixing of trajectories say III into II, i.e. a
wave packet localized on trajectory III at large $\xi$ and large $p$ (that is
at early times, $\eta$ small) has an amplitude $ \beta$ to be localized on
trajectory II at small $p$ ($\eta$ large). Note that had we taken $\omega <0$
rather than $\omega>0$, trajectory II would then have positive $p$ and
trajectory III negative $p$ rather than the reverse.

We now study  eq. (\ref{E3}) near the horizon where $ V \simeq - 1 + \xi / 4 M$.
In momentum space, after some elementary manipulation, the equation takes the
form 

\begin{equation}
\label{E4.3}
\left( {p^2 \over 4 M^2} \left [ \partial_p + {i \omega 4 M + 1 \over p} + i 4 M
\right ] ^2 + F^2 (p) \right ) \tilde \phi_\omega (p) = 0 \qquad .
\end{equation}

\noindent For $ \vert p \vert < 1 $ where $F(p) =p$, the solutions are

\begin{eqnarray}
\mbox{\rm I} &:& \tilde \phi_\omega (p) = \sqrt{ M \over \pi} \theta (\pm p) \vert
p\vert ^{-4 M i \omega -1} e^{-i 8 M p}\quad ; \quad \vert p \vert < 1 \quad ,\nonumber
\\ \mbox{\rm II, III} &:& \tilde \phi_\omega (p) 
= \sqrt{ M \over \pi} \theta (\pm p) \vert p \vert^{-4M i \omega
-1} \hspace{1.15 cm}\quad ; \quad \vert p \vert < 1 \quad .\label {E4.4}
\end{eqnarray}

\noindent Solutions in class I are associated with the trajectories of class I
($v$-modes) and are not involved in production (the singular
behavior of these modes is due to the presence of the (spurious) Cauchy horizon at $ \xi
= 8 M)$. For $ p \gg 1$,
where $F(p) = 1$, the solutions are

\begin{eqnarray}
\tilde \phi_\omega (p) = \sqrt{ M \over \pi} 
\theta (\pm p) \vert p \vert^{- 4
M i \omega - 1} e^{+i 4 M p} \vert p \vert^{i \alpha_\pm} \quad ; \quad  p \gg 1
\label{E4.5} \end{eqnarray}

\noindent where $\alpha _{\pm}$ are the roots of $ \alpha (\alpha - 1) - (4 M)^2 = 0$. We
take $ M \gg 1$ so that $ \alpha_\pm \simeq \pm 4 M - {i \over 2}$. These modes serve as
a basis of second quantization. Their positive and negative frequency parts will then
correspond to the bases of the quantized field wherein the annihilation part (positive
frequency) annihilates the in-vacuum (i.e. vacuum at large $ \xi$, hence large
$p$). To determine this we refer to the conserved scalar product which we norm
in the conventional way
\begin{eqnarray}
i\int\! d \xi\left[ \phi_{\omega^\prime}(\xi) ^* 
\left( \partial_\eta + V(\xi) \partial_\xi\right)
\phi_\omega (\xi)-\phi_{\omega^\prime}(\xi) ^*\leftrightarrow \phi_\omega (\xi)\right]
&=& \pm \delta (\omega - \omega^\prime) \nonumber\\
\label {E4.7a}
\end{eqnarray}
or in momentum space
\begin{eqnarray}
i 2\pi\int\! d p \left[ \tilde \phi_{\omega^\prime}(p) ^* 
\left( 
\partial_\eta - i p - {1/ 4M} - {p / 4 M} \partial_p 
\right)
\tilde \phi_\omega (p) \right.\!&&\nonumber \\
\left.-\tilde \phi_{\omega^\prime}(p) ^* \leftrightarrow 
\tilde \phi_\omega (p) \right]&=& \pm
\delta (\omega - \omega^\prime) \quad .\nonumber \\\label {E4.7}
\end{eqnarray}

\noindent Inserting  eq. (\ref{E4.5}) into  eq. (\ref {E4.7}) it is seen that the
sign of the scalar product is determined by the sign of $ \alpha_\pm $ in
 eq. (\ref {E4.5}).
The condition of positive frequency $(\alpha_+= +4M)$  corresponds to taking
wave packets localized along the trajectories $-p V(\xi) = - \omega + \vert
F(p) \vert$ rather than $-p V(\xi) = - \omega - \vert
F(p) \vert$. The additional requirement that the modes be localized along
trajectories II or III rather than I then imposes $p>0$. The condition of
positive $p$ which played an essential r{\^o}le in the preceding sections is
thereby recovered. In summary the in-modes appropriate for second quantization
which give rise to Hawking radiation are 
\begin{eqnarray} \tilde
\phi_\omega^{in} (p) = \sqrt{ M \over \pi}  \theta (p)  p^{- 4
M i \omega - 1} e^{-i 4 M p} p^{i 4M+1/2} \quad ; \quad  p \gg 1 \qquad .
\label{E4.8a} \end{eqnarray}

It remains to establish how
these in-modes which have been characterized at large $p$ evolve into their
forms  eq. (\ref {E4.4}) at small $p$.
To this end we use the WKB approximation whose validity we justify subsequently.
Thus we approximate the full solution of the in-mode for positive $p$ by

\begin{equation}
\label{E4.8}
\tilde \varphi_\omega^{in} (p) = \theta (p)  p^{- 4 M i \omega
-1} \; {e^{-i 4
M  [ p - \int^p d p \ F (p)  / p ]} \over \sqrt { F (p)  /
 p }} \qquad .
\end{equation}

\noindent Equation \ref{E4.8}) is exact for both $\vert  p \vert < 1$ and $
\vert  p \vert \gg 1$.
The validity of the WKB expansion is assured owing to the inequality 
\begin{equation} 
{ d(p/4M F(p)) \over dp }
 \ll
1 \qquad .
\label{E4.8aa}\end{equation}
This condition can be given a geometric interpretation. To this end
one reexpresses the solution $\xi(p)$ of the on mass shell condition
$H=0$ ( eq. (\ref{E4.1}) as $\xi(p) = \xi_F (p) + \Delta \xi(p)$ where
$\xi_F(p)$ corresponds to a point which moves with the fluid $\omega +
p V(\xi_F) =0$ and $\Delta \xi$ describes motion with respect to the
background $(p V^\prime (\xi_F) \Delta \xi )^2 - F^2(p)=0$. Then the
validity of the WKB approximation takes the form
\begin{equation} 
{d(1/\Delta \xi)\over dp}
 \ll
1 \qquad .
\label{E4.8b}\end{equation}
which expresses that the motion with respect to the background fluid
is sufficiently ``slow''.
The l.h.s. of this inequality is $O (M^{-1})$ which by assumption is $
\ll 1.$ Any back scattering due to the second order character of the
differential equation will be a non perturbative effect (typically of $
O (e^{-M}))$ which would result in the mixing of $u$ and $v$ type
modes. It is negligible.

Once having established that the low $p$ part of the in-modes is of the form
 eq. (\ref{E4.4}), types II and III, with only positive $p$, the
results of Section \ref{SS3} follow forthwith: to wit Fourier transform gives a
thermal distribution in $ \omega$ arising from packets built from the low
momentum part of the modes.

In any truncation scheme, be it linear or nonlinear, one may hope that
a condition of adiabaticity similar to  eq. (\ref{E4.8b}) will be
applicable. It will then imply that the creation of particles at a
Planckian distance from the horizon is strongly suppressed. Thus Unruh
vacuum will once more be a valid concept on scales $x>O(1)$ and the
usual spectrum of Hawking radiation will be recovered. Were there any
Planckian particles created at $x=O(1)$ they would severely disrupt the
Hawking flux. Their absence can be taken to be an expression of the
general principle (very well verified experimentally) that flat space
is stable against creation of Planckian particles. Since the curvature
in Schwarzschild geometry acts on scales $\Delta x =O(M)$, at a Planck
distance from the horizon space looks almost flat and the
principle can be applied with confidence. Thus the absence of created
Planckian particles at the horizon should come as no surprise.

\section{The classical trajectories}\label{SS6}

\figprov{8cm}{Ufiga}{The classical trajectories of outgoing null geodesics
(thin dotted curves) given by the Hamiltonian eq. \protect\ref{E18}) are
displayed in the Eddington-Finkelstein coordinate system. The light ray which
generates the horizon, the infalling shell, the origin $r=0$
and the singularity are also represented.}{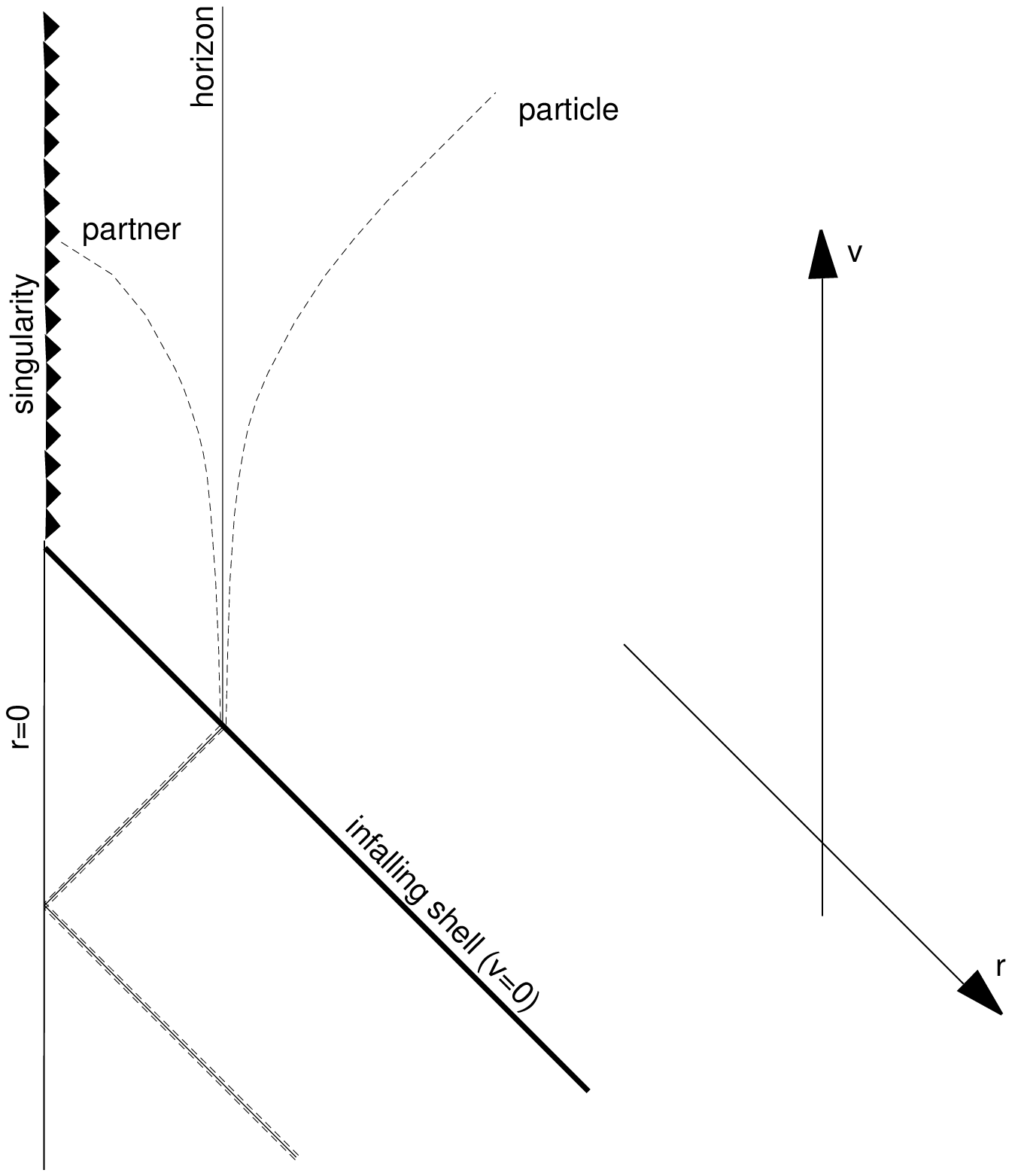}

We now have on hand three schemes on how to get Hawking radiation. It is
interesting to display in an Eddington-Finkelstein graph the classical trajectories corresponding
to the modes in the three cases. These encode vacuum fluctuations in the past
which are converted into quanta at $x= O(M)$. 

In Fig \ref{Ufiga} is displayed
the usual free field model. A vacuum fluctuation emerges from the star, it is a
pair that straddles the horizon. The fluctuation hugs the horizon at an
exponentially small distance. Outside the star it starts to propagate (outwards
for the observed Hawking photon and towards the singularity for its partner
-unobservable in Schwarzschild time but taking form on the other side of the
horizon in finite Kruskal time). Upon reaching $x=O(M)$ the fluctuation has
become an on mass shell quanta which now propagates along the trajectory $v-2r
=const$.

\figprov{8cm}{Ufigb}{The trajectories of outgoing light rays
given by the truncated Hamiltonian eq. \protect\ref{E23}) are displayed in the
Eddington-Finkelstein coordinate system. In this case the
trajectories no longer approach the horizon exponentially
but rather they stick at a planckian distance.}{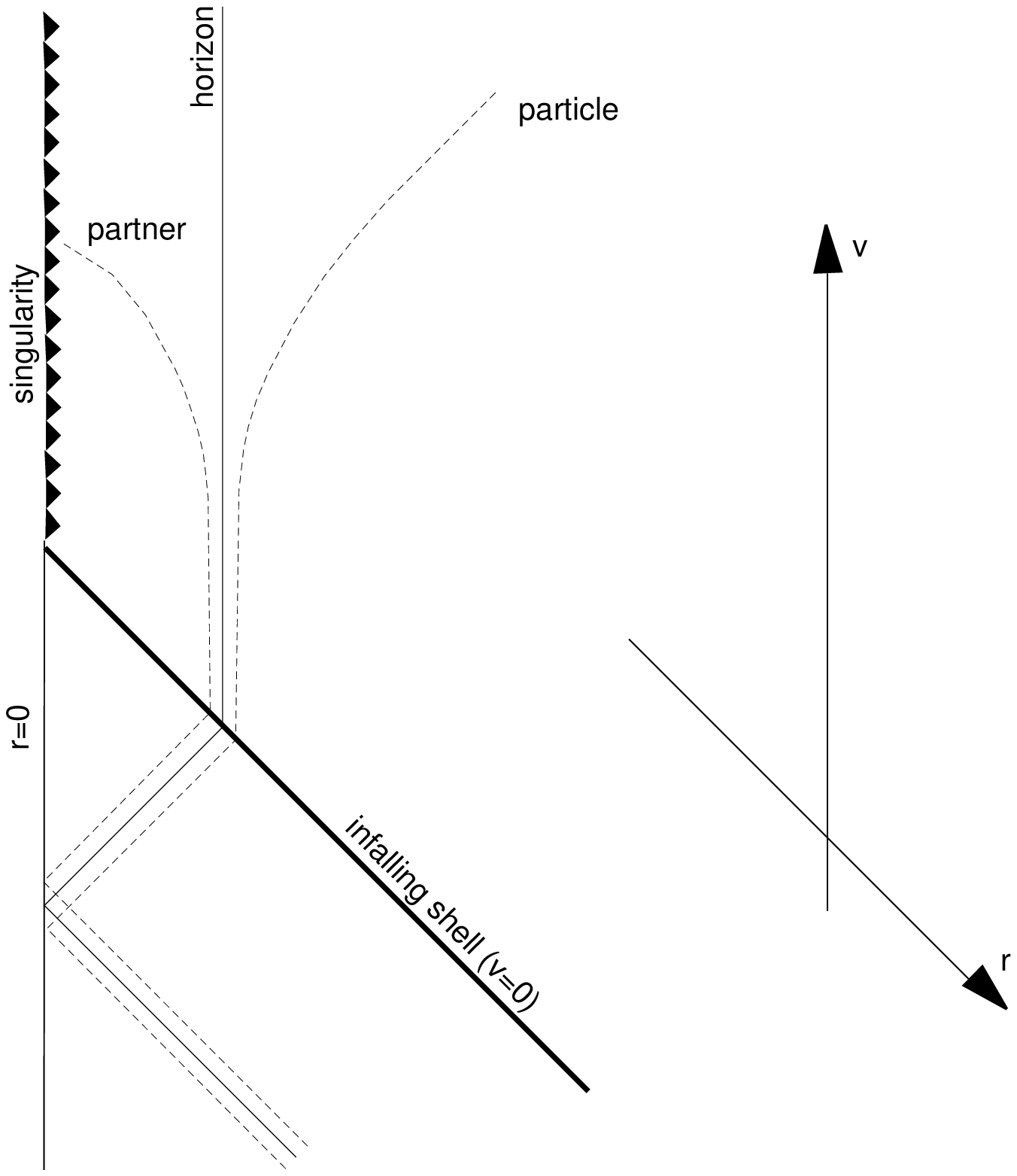}

In Fig \ref{Ufigb} the effect of the simple truncation of Section \ref{SS4}
is shown. We have taken $g(p)$ to be unity for $p>1$ (and of course
$g(p)=p$ for $p<1$). The propagation on either side of the horizon begins at
the edges of the Planckian zone $\vert x \vert =O( 1)$. The Planckian zone
$\vert x \vert <1$ is thus steadily solicited to give out radiation at a steady
rate.

\figprov{8cm}{Ufigc}{The  trajectories of  light rays 
in Unruh's truncated hydrodynamic model are displayed in the Eddington-Finkelstein coordinate
system. Only class II and III light rays have been displayed as it is these
which are responsible for production.  Note that since $v$
and $\eta$ differ by a regular function of $x$, the
trajectories in the coordinate system $\eta , \xi$ given by the
Hamiltonian eq. \protect\ref{E4.1}) would be very similar to those depicted in
the figure.}{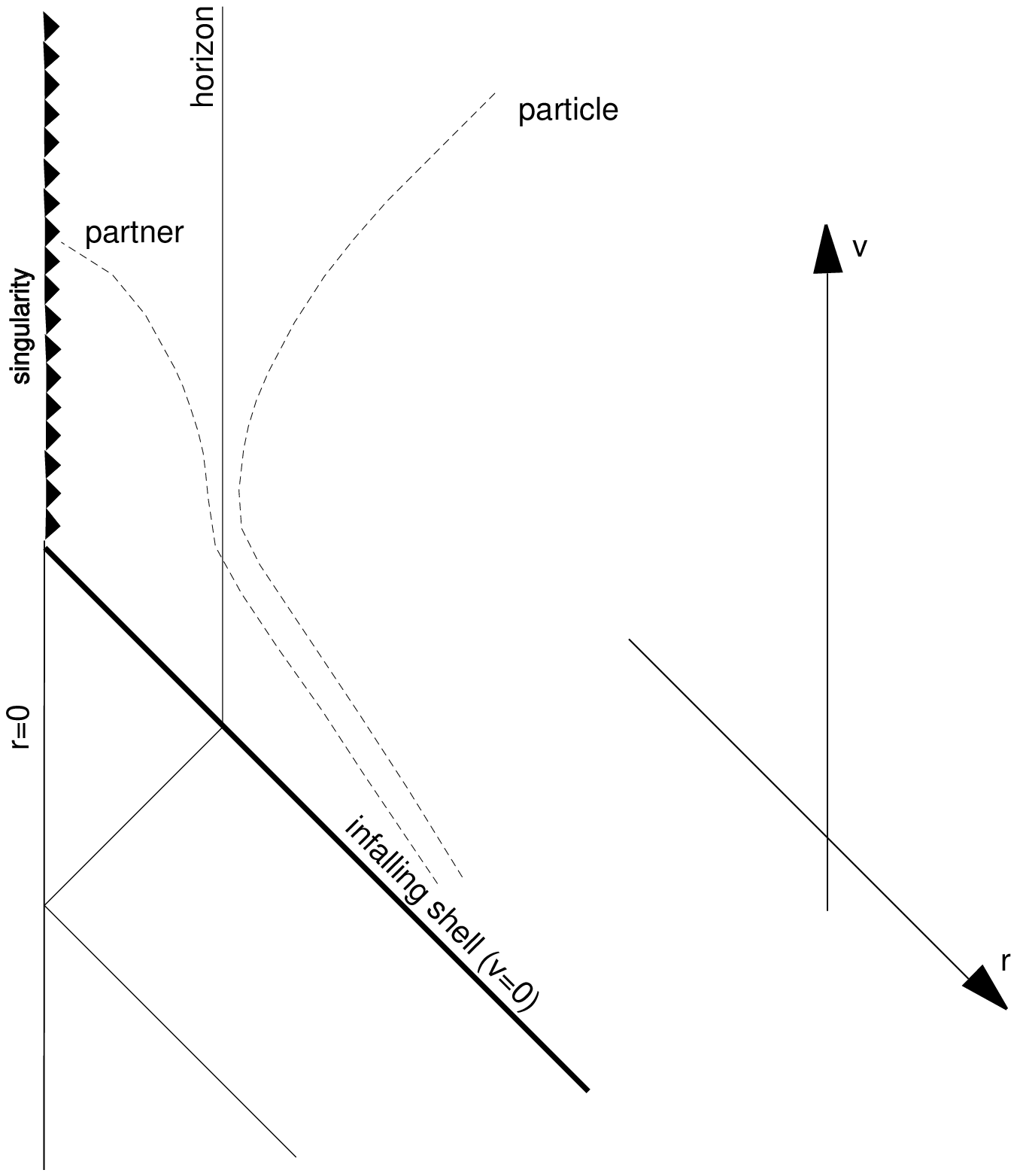}

Finally in Fig. \ref{Ufigc} the trajectories in Unruh's truncation are
sketched. Here they begin from ${\cal I}^-$, but unlike the free field case
they do not go into the star to reflect and then come out as in Fig
\ref{Ufiga}. Rather they reach the horizon region directly whereupon at 
$ x =\pm 1$ they reflect. The amplitude of the reflected wave is augmented; it
is accompanied by a partner which appears on the other side with amplitude
$\beta$ and the production is encoded in $\alpha^2 - 1 = \beta^2 = (e^{\beta
\omega} -1 )^{-1}$. It is fairly difficult to interpret this structure in terms
of the black hole. In the fluid it comes about due to the ``riding in'' of the
modes on the background flow.
\par More important, from the figures it is seen that in all cases, at
large distance from the horizon $\vert x \vert \gg 1$, $\vert p \vert
\ll 1$, one recovers the usual free particle trajectories. Therefore,
independently of the truncation we have used, Hawking radiation remains
a pair production phenomenon: the outgoing quanta are accompagned by
partners on the other side of the horizon.

\section{Conclusion}\label{SS7}
It appears to us that the scenario based on  eq. (\ref{E21a}) could well turn
out to reflect some part of the truth. The function $g(p)$ deviates from its free field
value, $p$, for $p>1$ so as to give rise to a sort of quasi-particle
description. The strong gravitational interaction among the modes will result
in continuous mixing of angular momenta, so in the rigorous theory,
restriction to s-waves will no longer be possible. The whole medium is to be
regard as a matter gravitational soup, which has some s-wave content. For
values of $x$ greater than unity this content becomes that of the usual free
field and there should be a turn over from a quasi to true particle theory.
The transplanckian soup steadily feeds into the free field
sector. It should be possible to display the transition region ($x=O(1)$,
$p=O(1)$) by perturbative methods, for example by expanding the gravitational
action to quadratic terms in fluctuations around the background geometry. This
will show how the modes start interacting with each other as they move into the
Planckian region. We might expect $g(p)$ to become complex for $p>1$,
corresponding to a Planckian lifetime of the s-wave quasi-particle. Such
behavior might encode that fact that an s-wave gets swallowed up in the
extrapolation backwards in time towards the Planckian skin. Then instead of the
extrapolation through the star drawn in Fig \ref{Ufigb}, the modes just peter
out within the skin. This is indicated by some shading in Fig \ref{Ufigd}.

\figprov{8cm}{Ufigd}{The  trajectories of  light rays 
 in the Eddington-Finkelstein coordinate
system if the truncation $g(p)$ where complex. In this case
the trajectories disapear into some quantum fuzz which is
represented by some shading in the figure.}{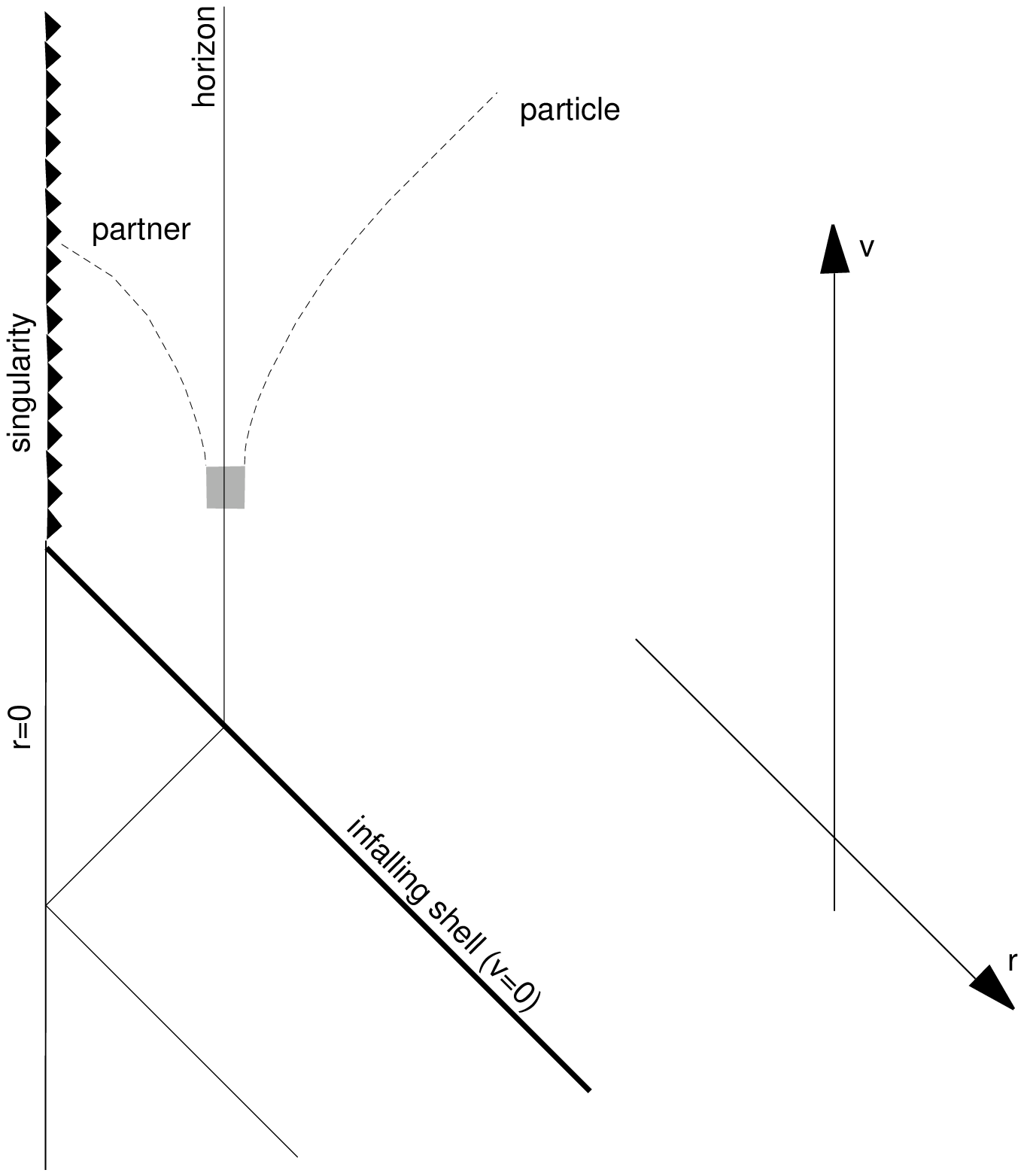}

In addition there is the interaction of the modes with the gravitational field
emanating from the sources which constitute the star (possibly taken together
with the mean effect due to the other modes, as in the semiclassical theory).
Here it would be ``recoil'' effects of the gravitational field which would be
responsible for deviations from free field theory. That such effects can
be important has recently been shown by one of us (RP) in the context of
the accelerated detector \cite{Par}, and more recently corroborated in a study of
accelerating mirrors \cite{ParII}. What the incidence of these two types of
effects on the modes is for the unitarity problem 
 remains to be seen.

Finally one must be prepared to encounter the very strong coupling problem
which will arise well into the transplanckian region where 
conceptual problems will arise  concerning the nature of space--time. 

Whatever, on the basis of the above considerations, we conjecture strongly that
Hawking radiation is protected from the vicissitudes of quantum gravity. It
appears as an essentially kinematic response to the presence of the event
horizon encountered in the collapse of a macroscopic black hole.

\vskip 1cm
 \noindent{\bf Added note}
After this manuscript was completed, C. Bouchiat and F.
Englert called our attention to the following conceptual problem.
The truncation we have used treats u-modes and v-modes
asymmetrically, thereby explicitly breaking the invariance of the theory
under local Lorentz transformations. However one should recall that the
formation of a black hole by the collapse of a star induces an
asymmetry between u-ness and v-ness:  only a future
horizon is formed but no past horizon exists. Concommitantly the field
state is asymmetric, the v-part being Schwarzschild (Boulware vacuum)
in character whereas
the u-part is Kruskal (Hartle-Hawking)
in character (i.e. the state is Unruh vacuum).

We beleive that it is  not unreasonable to envisage that a phenomenological truncation
of the field equations is state dependent since it should encode the
dynamics of the matter field state induced by the quantum gravitational
interactions. In Unruh vacuum the truncation would then treat u and v
modes differently. What would be the effect of a symmetric truncation
in a symmetric state such as the eternal black hole situation
remains to be seen.

\end{document}